\documentstyle[prd,aps,eqsecnum,epsf,preprint]{revtex}

\draft
\begin{document}
\thispagestyle{empty}
{\baselineskip0pt
\leftline{\baselineskip16pt\sl\vbox to0pt{\hbox
{\it Department of Physics}
               \hbox{\it Waseda University}\vss}}
\rightline{\baselineskip16pt\rm\vbox to20pt{\hbox{WU-AP/133/01}
            \hbox{\today} 
\vss}}%
}
\vskip1cm
\begin{center}{\large \bf
Critical phenomena in Newtonian gravity}
\end{center}
\vskip1cm
\begin{center}
 {\large 
Hideki Maeda
\footnote{Electronic address: hideki@gravity.phys.waseda.ac.jp}
and Tomohiro Harada
\footnote{Electronic address: harada@gravity.phys.waseda.ac.jp}\\
{\em Department of Physics,~Waseda University, Shinjuku, 
Tokyo 169-8555, Japan}}
\end{center}

\begin{abstract}
We investigate 
the stability of self-similar solutions for a gravitationally collapsing isothermal sphere in Newtonian gravity by means of a normal mode analysis. It is 
found that the Hunter series of solutions are 
highly unstable, while
neither the Larson-Penston 
solution nor the homogeneous collapse 
one have an analytic unstable mode.
Since the homogeneous collapse solution is known 
to suffer the kink instability,
the present result and recent numerical simulations strongly support 
a proposition that the Larson-Penston solution will be realized in 
astrophysical situations. 
It is also found that the Hunter (A) solution has a single unstable mode,
which implies that it is a critical solution associated with 
some critical phenomena
which are analogous to those in general relativity.
The critical exponent $\gamma$ is calculated as $\gamma\simeq 0.10567$.  
In contrast to the general relativistic case, the order parameter will be 
the collapsed mass. In order to obtain a complete picture of the Newtonian 
critical phenomena,
full numerical simulations will be needed.
\end{abstract}


\section{INTRODUCTION}
Spherically symmetric self-similar systems have been widely studied in the context of both Newtonian gravity and general relativity. Self-similar solutions in Newtonian gravity have been studied in an effort to obtain realistic solutions of gravitational collapse leading to star formation~\cite{penston1969,larson1969,shu1977,hunter1977}. In this context, 
Larson and Penston independently found a self-similar solution, which 
describes a gravitationally collapsing isothermal gas 
sphere~\cite{penston1969,larson1969}. Thereafter, 
Hunter found a new series of self-similar solutions, 
and that a set of such solutions is infinite and discrete
~\cite{hunter1977}. Whitworth and Summers investigated 
the mathematical structure of the equation for 
self-similar solutions in more detail
and found a band structure of a new family of 
self-similar solutions with loss of analyticity~\cite{WS1985}. 
The solutions were classified into 
two types, based on the behavior around the sonic point.
These self-similar solutions were generalized to general relativity
by Ori and Piran~\cite{op1987,op1990}.

In general relativity, self-similar solutions called attention
in the discovery of the critical behavior by Choptuik~\cite{choptuik1993}.
Evans and Coleman found similar critical behavior 
in the collapse of a radiation fluid~\cite{ec1994}.
A renormalization group approach showed that the critical solution,
which is at the threshold of the collapse to a black hole,
is an analytic self-similar solution with the unique unstable 
mode and that
the critical exponent which appears 
in the scaling law of the formed black hole mass 
is equal to the inverse of the eigenvalue of 
the unstable mode for a perfect fluid case~\cite{kha1995,maison1996,kha1999}. 
In the previous work~\cite{hm2001}, the authors found 
that the general relativistic counterpart of the Hunter (A) solution
has a single unstable mode and expected that the Hunter (A) solution
is a critical solution of the Newtonian counterpart of the critical 
behavior in gravitational collapse. 
The first purpose of this paper is to confirm this expectation
and to 
calculate the accurate value of the critical exponent.

In addition to this critical nature,
we should mention the role of a self-similar solution as an attractor
of gravitational collapse.
Recent numerical simulations and results of mode analyses 
showed that the Larson-Penston solution is the best 
description for the central part of a collapsing gas 
sphere in Newtonian gravity~\cite{ti1999,hn1997,fc1993,hm2000a,hm2000b} 
and in general relativity~\cite{hm2001,harada1998}. 
Hanawa and Nakayama~\cite{hn1997} examined the spherically symmetric 
unstable modes in linear order of which the growth is
faster than $|t|^{-1}$ as $t\to 0$. 
They showed that the Larson-Penston solution has 
no such mode, while the Hunter (B) and (D) solutions 
have some unstable modes by means of a normal mode analysis.
They concluded that the Hunter (B) and (D) solutions 
are unstable and not likely to be realized in generic situation, 
while the Hunter (A) and (C) solutions were dropped from the analysis.
There is no reason
to rule out the Hunter (A) and (C) solutions {\em a priori} .
The second purpose of this paper is to
complete the stability analysis on the self-similar solutions 
including these solutions.

The organization of this paper is the following. In section II, 
basic equations are presented. In section III, 
eigenvalues of unstable modes for 
the self-similar solutions are presented. 
Section IV is devoted to discussions.
In Section V, we summarize the paper.

\section{BASIC EQUATIONS}
\subsection{Basic equations in the zooming coordinates}
A gravitationally collapsing isothermal sphere is described in spherical coordinates by 

\begin{eqnarray}
&&\frac{\partial\rho}{\partial t}+\frac{1}{r^2}\frac{\partial}{\partial r}(r^2\rho v)=0, \label{b1}\\
&&\frac{\partial}{\partial t}(\rho v)+\frac{1}{r^2}\frac{\partial}{\partial r}(r^2\rho v^2)+c_s^2\frac{\partial\rho}{\partial r}+\rho\frac{G M}{r^2}=0,  \label{b2}\\
&&\frac{\partial M}{\partial t}+v\frac{\partial M}{\partial r}=0,  \label{b3}\\
&&\frac{\partial M}{\partial r}=4\pi r^2 \rho, \label{b4}
\end{eqnarray}
where $\rho,v,M,c_s$ and $G$ denote the density, radial velocity, total mass inside the radial coordinate $r$, sound speed, and gravitational constant, respectively. We introduce the zooming coordinate $z=(c_s t)/r$.
We also introduce dimensionless functions $U,P$ and $m$:

\begin{eqnarray}
v(r,t)&=&-c_s U(r,t), \\
\rho(r,t)&=&\frac{c_s^2 P(r,t)}{4\pi G r^2} ,\\
M(r,t)&=&\frac{c_s^3 t m(r,t)}{G}.
\end{eqnarray}
Then, the equations are expressed as

\begin{eqnarray}
&&\frac{t}{z}\dot P+(1+zU)P'+zPU'=0, \label{basic1} \\
&&-t(\dot U P+ U\dot P)-z(U+zU^2+z)P' \nonumber \\ 
&&-zP(1+2zU)U'-2zP+mPz^2=0,\label{basic2} \\
&&t\dot m+zm'+m-PU=0,\label{basic3} \\
&&-z^2m'=P, \label{basic4}
\end{eqnarray}
where the prime and dot denote the partial derivatives 
with respect to $z$ and $t$, respectively. 

\subsection{Self-similar solutions}
\begin{figure}[htb]
\centerline{
\epsfxsize 18cm \epsfysize 20cm
\epsfbox{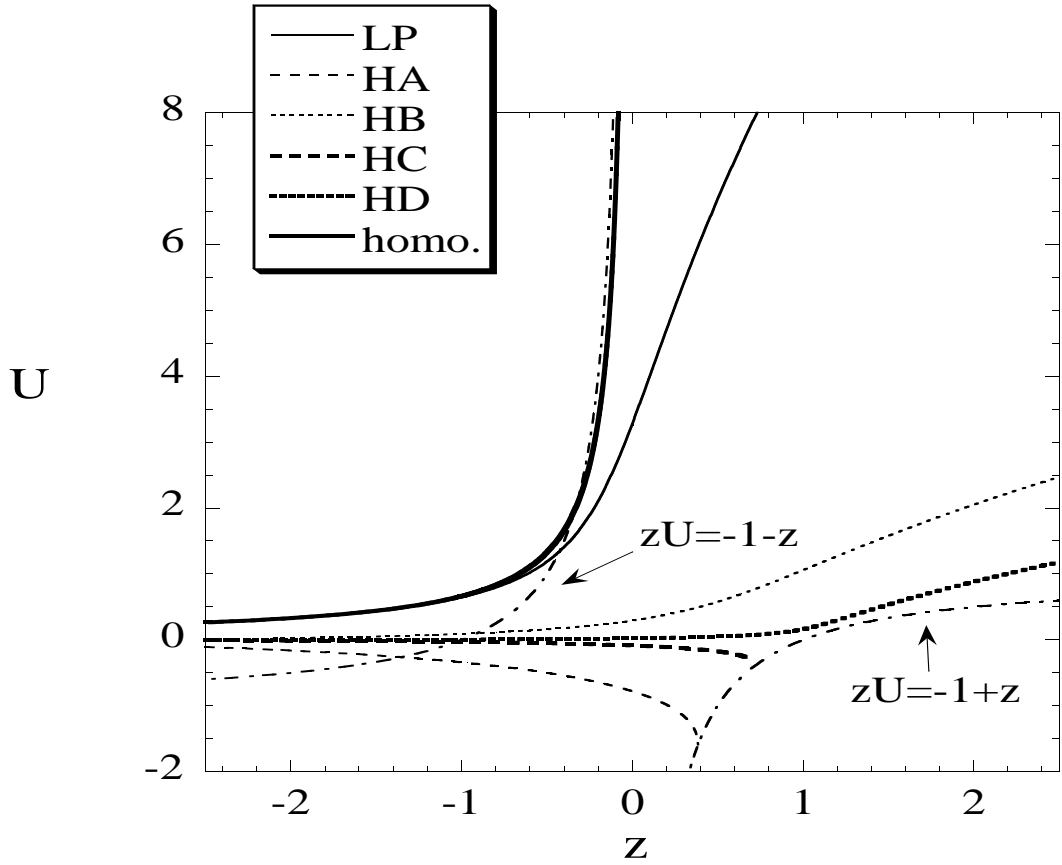}}
\caption{$U=-v/c_s$ for self-similar solutions are plotted. The Hunter (A) and (C) solutions terminate at the second sonic point $zU=-1+z$.}
\label{U}
\end{figure}

\begin{figure}[htb]
\centerline{
\epsfxsize 18cm \epsfysize 20cm
\epsfbox{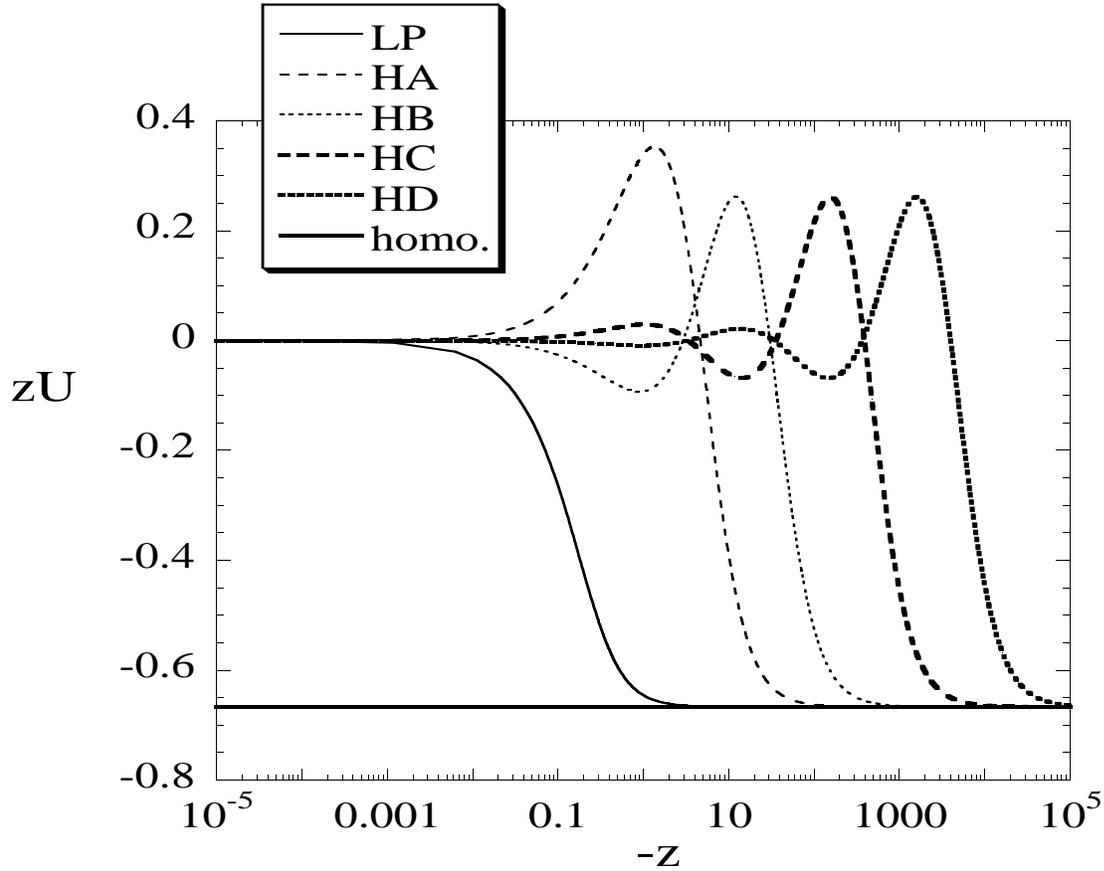}}
\caption{$zU=-zv/c_s$ for self-similar solutions are plotted for $z<0$. $-z \to \infty$ corresponds to the center. The Larson-Penston solution has no node which means that it is a pure collapse solution. Some oscillations can be seen in the Hunter (A)-(D) solutions. The number of oscillations is one, two, three and four for the Hunter (A), (B), (C) and (D) solution, respectively.}
\label{zU}
\end{figure}

\begin{figure}[htb]
\centerline{
\epsfxsize 18cm \epsfysize 20cm
\epsfbox{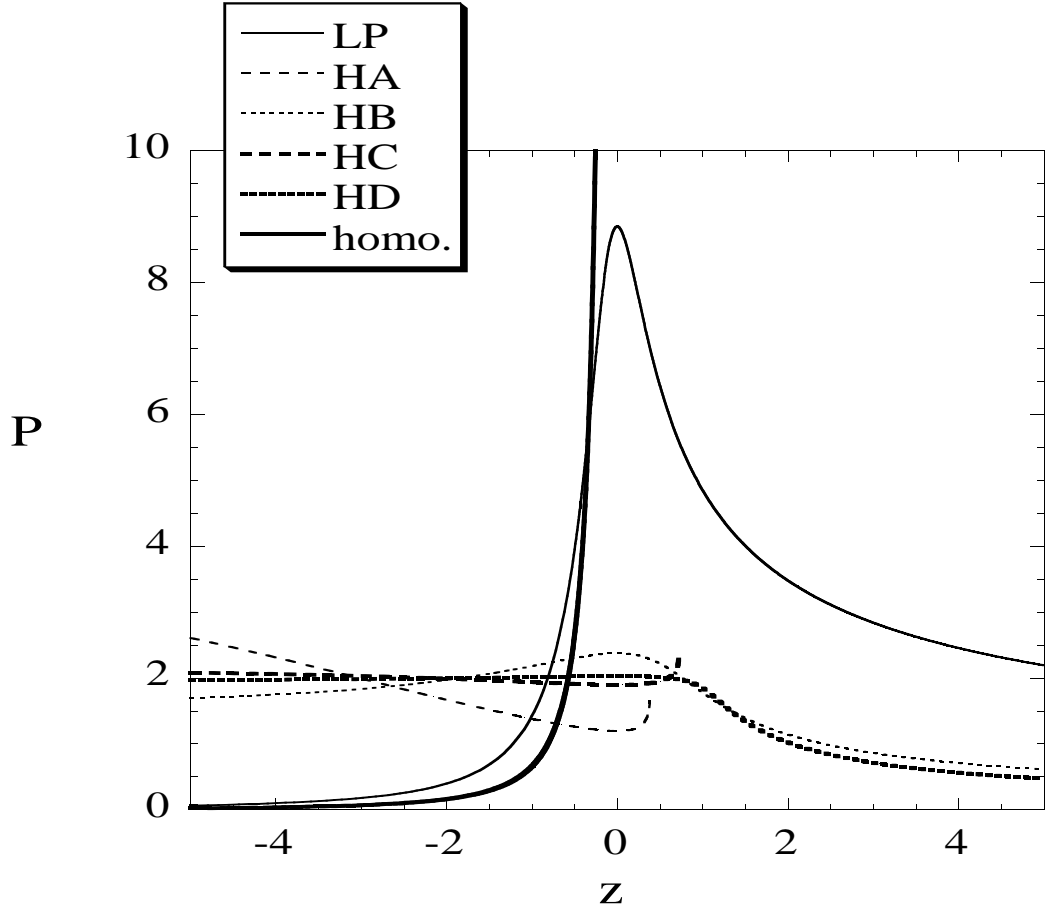}}
\caption{$P=4\pi G r^2 \rho/c_s^2$ for self-similar solutions are plotted. The Hunter (A) and (C) solutions terminate at the second sonic point and their density are finite there. The homogeneous collapse solution has a big-crunch singularity at $z=0$. For the Larson-Penston solution, the Hunter (A) and (C) solutions, the density is diluted in the $z>0(t>0)$ evolution.}
\label{P}
\end{figure}

\begin{figure}[htb]
\centerline{
\epsfxsize 18cm \epsfysize 20cm
\epsfbox{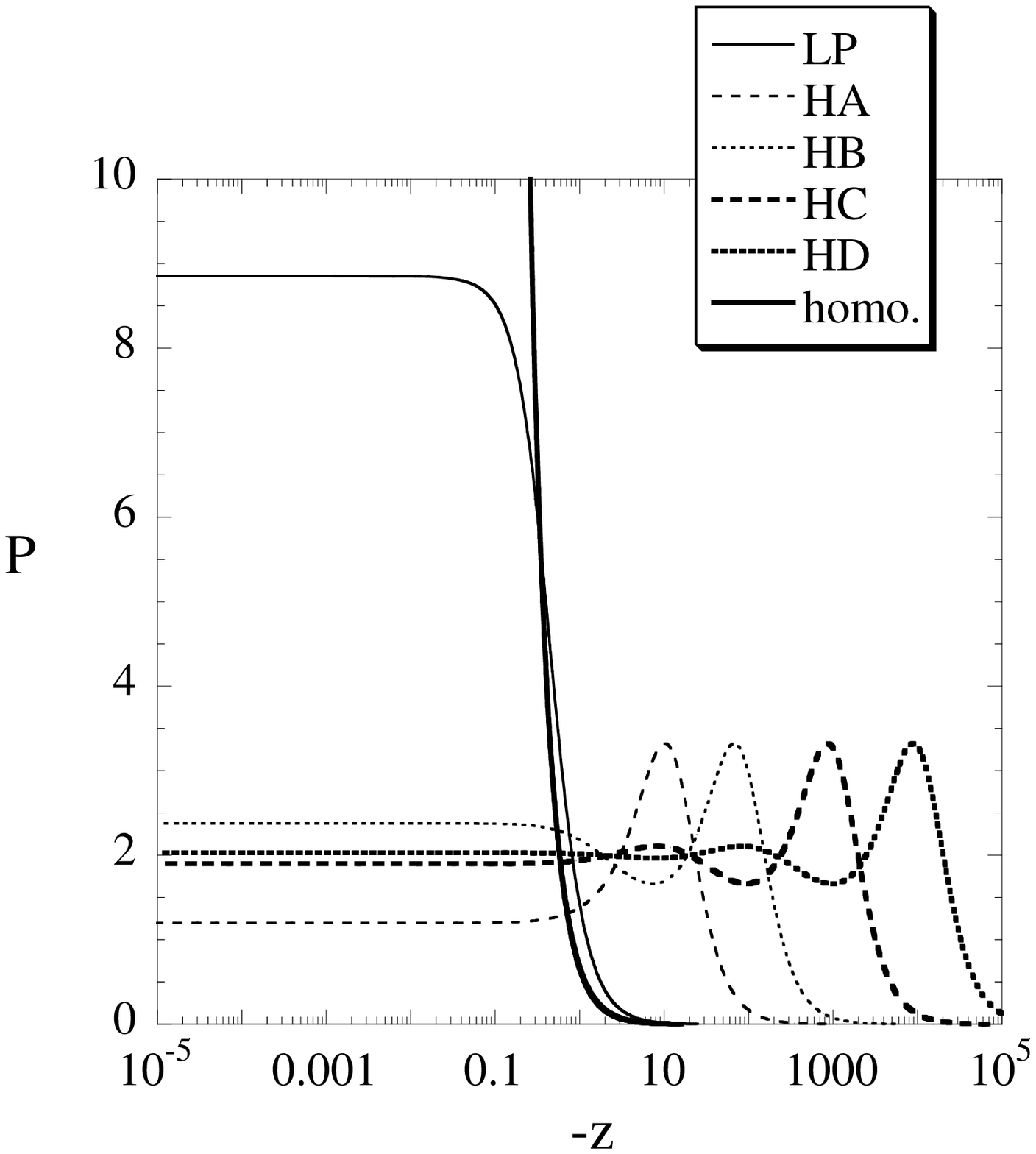}}
\caption{$P=4\pi G r^2 \rho/c_s^2$ for self-similar solutions are plotted for $z<0$.  $-z \to \infty$ corresponds to the center. The homogeneous collapse solution has a big-crunch singularity at $z=0$. Some oscillations can be seen in the Hunter (A)-(D) solutions. The number of oscillations is one, two, three and four for the Hunter (A), (B), (C) and (D) solution, respectively. No oscillation can be seen in the the Larson-Penston solution.}
\label{Plarge}
\end{figure}

\begin{figure}[htb]
\centerline{
\epsfxsize 18cm \epsfysize 20cm
\epsfbox{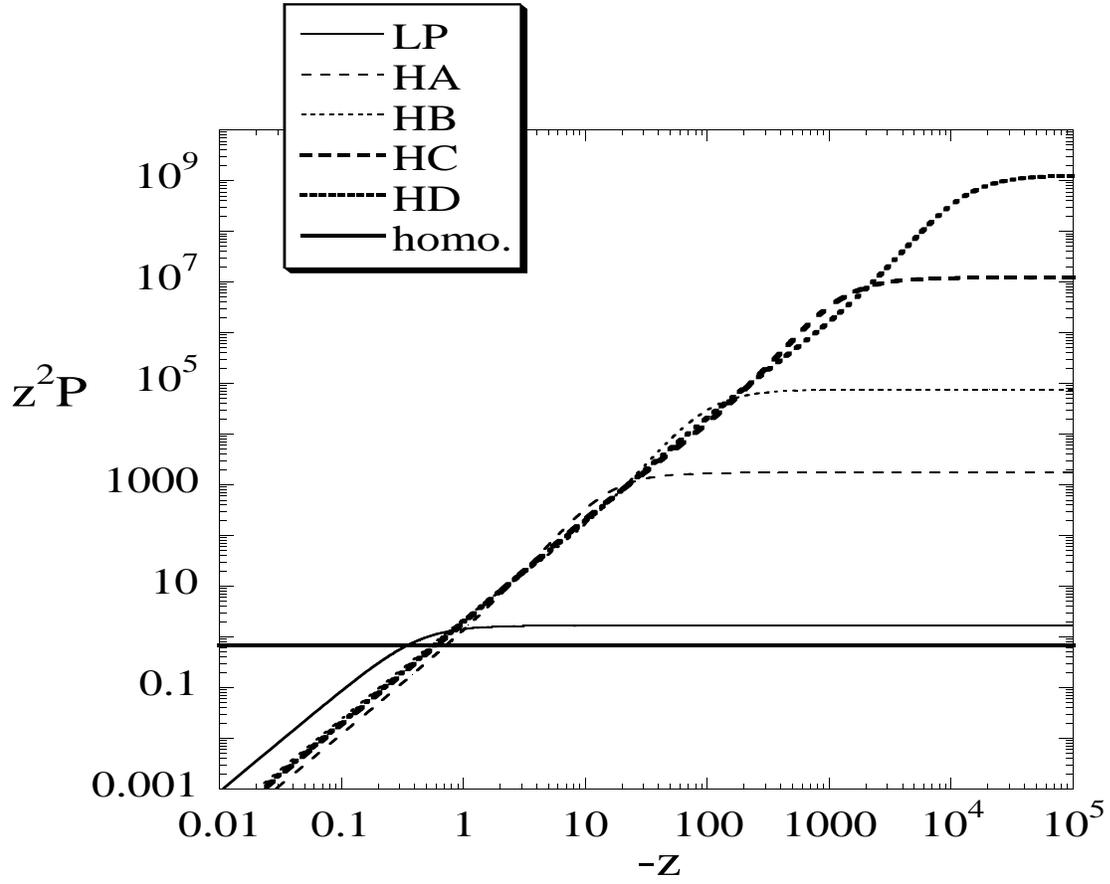}}
\caption{$z^2P=4\pi G t^2 \rho$ for self-similar solutions are plotted for $z<0$. The Hunter (D) solution has the highest central value among these solutions.}
\label{z^2P}
\end{figure}

Self-similar solutions are characterized by $U=U(z), P=P(z)$ and $m=m(z)$. From this ansatz, equations (\ref{basic1})-(\ref{basic4}) become 

\begin{eqnarray}
U'&=&\frac{(zU+1)[P(zU+1)-2]}{(zU+1)^2-z^2}, \\
P'&=&\frac{zP[2-P(zU+1)]}{(zU+1)^2-z^2}, \\
m&=&P(U+1/z).
\end{eqnarray}
If the analyticity at the center is assumed, self-similar solutions can be expanded in Taylor series around $z \to -\infty$, i.e., $t \to -\infty$
or $r \to 0$ as follows:

\begin{eqnarray}
&&U=-\frac{2}{3z}-\frac{1}{45}\left(\frac23-e^{Q_0}\right)\frac{1}{z^3}+O\left(\frac{1}{z^5}\right), \nonumber \\
&&Q=\ln(z^2P)=Q_0+\frac16\left(\frac23-e^{Q_0}\right)\frac{1}{z^2}+O\left(\frac{1}{z^4}\right).
\end{eqnarray}
Therefore, self-similar solutions 
which have regular center are specified by one parameter
$Q_0$. Taylor series expandability 
in the neighborhood of the sonic point $z=z_s$ ($zU+z=-1$) requires that there are two possible analytic solutions. The first solution (type 1) is 
\begin{eqnarray}
&&U=\left(1+\frac{1}{z_s}\right)\left[-1+\frac{z-z_s}{z_s}-\frac{(z-z_s)^2}{2z_s^2} \cdots\right], \nonumber \\
&&P=-\frac{2}{z_s}-\frac{2(1+z_s)(z-z_s)}{z_s^3}-\frac{(1+z_s)^2(z-z_s)^2}{z_s^5} \cdots ,
\end{eqnarray}
and the second one (type 2) is 
\begin{eqnarray}
&&U=-\left(1+\frac{1}{z_s}\right)-\frac{z-z_s}{z_s}+\frac{(z_s^2-z_s-1)(z-z_s)^2}{2z_s^3(3z_s+2)}\cdots, \nonumber \\
&&P=-\frac{2}{z_s}+\frac{2(z-z_s)}{z_s^2}-\frac{(7z_s^2+6z_s+1)(z-z_s)^2}{z_s^4(3z_s+2)}\cdots.
\end{eqnarray}
Therefore, self-similar solutions which are analytic at the sonic point are specified by one parameter $z_s$ around the sonic point. We can find one exact solution and numerical ones with analyticity both at the center 
and at the sonic point. The former is a homogeneous collapse solution:
\begin{equation}
Q_0=\ln{\frac23},\quad z_s=-\frac{1}{3}, \quad P=\frac{2}{3z^2},\quad U=-\frac{2}{3z},\quad m=\frac{2}{9z^3}.
\end{equation} 
The values of $Q_0$ and $z_s$ for numerical solutions (the Larson-Penston solution and the Hunter (A)-(D) solutions) are summarized in 
Table \ref{table1}. These self-similar solutions 
are displayed in Figs. \ref{U}-\ref{z^2P}. 

The homogeneous collapse solutions is the
only solution which has the big crunch singularity. The big crunch occurs at $t=0$, i.e., the singularity occurs at the same time everywhere. Unlike the homogeneous solution, the Larson-Penston solution and the Hunter (A)-(D) solutions are regular at $t=0$, except for at $r=0$. The Hunter (A) and (C) 
solutions encounter another sonic point.
Neither the Hunter (A) nor (C) solutions can pass through the 
second sonic point regularly (see Figs. \ref{U} and \ref{P}). The Hunter (A)-(D) solutions are characterized by the number of oscillations in their profiles $zU$ (see Fig. \ref{zU}). Unlike the Hunter series, 
no oscillation can be seen in the profile for 
the Larson-Penston solution. This means that the Larson-Penston solution is a pure collapse solution. A similar property is shown also in $P$ (see Fig. \ref{Plarge}). In the Larson-Penston solution and the Hunter (A)-(D) solutions, the density profile has 
the maximum value at the center and decreases monotonically with increasing in $z$ (see Fig. \ref{z^2P}). In these solutions, the Hunter (D) solution has the biggest central value of $P$. We note that 
the Hunter solutions and the homogeneous collapse one are type 1, 
while the Larson-Penston solution is type 2.

\subsection{Perturbation equations}
We consider the spherically symmetric linear perturbations around the self-similar solution. We define the perturbation quantities as

\begin{eqnarray}
P(r,t)&=&P_0(z)+\varepsilon P_1(t,z)+O(\varepsilon^2), \nonumber \\
U(r,t)&=&U_0(z)+\varepsilon U_1(t,z)+O(\varepsilon^2), \nonumber \\
m(r,t)&=&m_0(z)+\varepsilon m_1(t,z)+O(\varepsilon^2),
\end{eqnarray}
where $P_0,U_0$ and $m_0$ are the background self-similar solution and $\varepsilon$ is a small parameter which controls the expansion. Then we find the equations for perturbations up to linear order of $\varepsilon$ as

\begin{eqnarray}
&&\frac{t \dot{P_1}}{z}+z U_1 P'+P_1'(1+z U)+z P U_1'+z P_1 U'=0, \\
&&-z P'(U_1+2z U U_1)-zP_1'(U+z U^2+z) \nonumber \\
&&- t \dot{U_1}P- t U \dot{P_1}-z(P_1 U'+U_1' P)(1+2z U) \nonumber \\
&&-2z^2 U_1 P U'-2zP_1+z^2(P m_1+m P_1)=0, \\
&&z m_1'+m_1+t \dot{m_1}-(UP_1+PU_1)=0, \\
&&-z^2 m_1'=P_1,
\end{eqnarray}
where we have omitted the suffix $_0$ for simplicity.

We assume the time dependence of the perturbations as 

\begin{eqnarray}
P_1(r,t)=\delta P(z) e^{\sigma \tau}, \nonumber \\
U_1(r,t)=\delta U(z) e^{\sigma \tau}, \nonumber \\
m_1(r,t)=\delta m(z) e^{\sigma \tau},\label{tdepend}
\end{eqnarray}
where $\tau \equiv -\ln(-t)$. Then we find the following equations for the 
perturbations:

\begin{eqnarray}
&&[(1+zU)^2-z^2]\delta P' \nonumber \\
&=& \left[2z-Pz(1+zU)+\frac{\sigma}{z}(1+zU)-\frac{zP}{1-\sigma}(1+zU)\right]\delta P \nonumber \\
&&+\left[-\sigma P+z^2PU'-(1+zU)zP'-\frac{z^2P^2}{1-\sigma}\right]\delta U, \label{peq1} \\
&&-zP[(1+zU)^2-z^2]\delta U' \nonumber \\
&=&z\left[(1+zU)^2\left(U'-P-\frac{P}{1-\sigma}\right)+2(1+zU)+\sigma-z^2U'\right]\delta P \nonumber \\
&&+\left[(1+zU)\left(-\sigma P+z^2PU'-\frac{z^2P^2}{1-\sigma}\right)-z^3P'\right]\delta U, \label{peq2} \\
&&-z^2 \delta m'=\delta P,  \label{peq4} \\
&&(1-\sigma)\delta m=\left(\frac{1}{z}+U\right)\delta P+ P\delta U. \label{peq3} 
\end{eqnarray}

Here we examine boundary conditions which the perturbations should satisfy at the boundaries. First we consider the regular center ($z=-\infty$). The perturbations near the center must satisfy

\begin{eqnarray}
&&\delta U=\frac{\delta U_0}{z}, \nonumber \\
&&\delta P=-\frac{3e^{Q_0}\delta U_0}{\sigma z^2}.
\end{eqnarray}
Next we consider the sonic point ($z=z_s$). We require that the perturbation of the density gradient is finite. The boundary condition for the perturbations at the sonic point is 
\begin{equation}
-\left[(1-\sigma)\left(P+\frac{2}{z}-\frac{\sigma}{z^2}\right)+P\right]\delta P+\left[\frac{(1-\sigma)(-z^2 P'+\sigma P-z^2 PU')}{z^2}+P^2\right]\delta U=0.
\end{equation}
Only for a discrete set of $\sigma$, there exists a solution of perturbation equations which is regular both at the regular center and at the sonic point. Thus we can obtain eigenvalues $\sigma$ and the associated eigenmodes. It is easily shown that the homogeneous collapse solution has one stable mode $\sigma=-2/3$ (see Appendix A). We note that all self-similar solutions have 
a ghost mode with $\sigma=1$ (see Appendix B).

\section{NUMERICAL RESULTS}
We have solved equations (\ref{peq1}) and (\ref{peq2}) numerically by fourth-order Runge-Kutta-Gill integration with adaptive stepsize control and obtained eigenvalues of them by the shooting method. We have set $\delta U_0=1$. When the amplitude of the perturbations has grown up to larger than $10^{10}$, we have appropriately scaled down their amplitude equally in order for the calculation not to be stopped by overflowing error. We have used equation (\ref{peq3}) to check the numerical error. The numerical error of this calculation has been within $10^{-7}$. We have assumed that the eigenvalue is positive. Since $\sigma$ has upper bound $\sqrt{\exp(Q_0)}+1$ for the self-similar solutions~\cite{hn1997}, we have sought the eigenvalues in the region $0<\sigma<\sqrt{\exp(Q_0)}+1$. 
The result is summarized in Table \ref{table2}. 

\begin{table}[htbp]
	\begin{center}
		\begin{tabular}{lccc} 
		Solution & $Q_0$ & $\sqrt{\exp(Q_0)}+1$ & $z_s$  \\ \hline\hline
		Homogeneous & $\ln{(2/3)}$ & $1.8165$ & -1/3  \\ \hline
		Larson-Penston & 0.5101 & $2.2905$ & -0.4271  \\ \hline
		Hunter (A) & 7.45616 & $42.599$ & -1.35305  \\ \hline
		Hunter (B) & 11.236 & $276.34$ & -0.9071  \\ \hline
		Hunter (C) & 16.322 & $3502.7$ &  -1.0295  \\ \hline
		Hunter (D) & 20.975 & $35865$ & -0.9911  \\ 
		\end{tabular}
	\end{center}
	\caption{$Q_0$ and $z_s$ of the self-similar solutions.}
	\label{table1}
\end{table}

\begin{table}[htbp]
	\begin{center}
		\begin{tabular}{lcc} 
		Solution &  Mode & $\sigma$ \\ \hline\hline
		Homogeneous & & Nothing \\ \hline
		Larson-Penston &  & Nothing \\ \hline
		Hunter (A) &  1 & 9.4637 \\ \hline
		Hunter (B) &  1 & $ 5.49 $ \\
		          & 2 & $5.74 \times 10^1$ \\ \hline
		Hunter (C) &  1 & 6.56 \\
		          & 2 & $5.89 \times 10^1$ \\
		          & 3 & $7.18 \times 10^2$ \\ \hline
		Hunter (D) &  1 & 6.22 \\ 
		          & 2 & $5.85 \times 10^1$ \\
		          & 3 & $5.95 \times 10^2$ \\
		          & 4 & $7.35 \times 10^3$ \\ 
		\end{tabular}
	\end{center}
	\caption{Summary of stability analysis (unstable modes).}
	\label{table2}
\end{table}

It is found that neither the homogeneous collapse solution nor 
the Larson-Penston one have 
no unstable mode and that each Hunter solution has the same number of unstable modes as the number of oscillations in their profiles of $zU$. We obtain the eigenvalues up to five digits only for the Hunter (A) solution since this relates to the critical exponent in critical phenomena. We number the unstable modes for each Hunter solution in order of magnitude of their eigenvalue. The eigenvalues of mode $n (n=1,2,3)$ for the Hunter (B), (C) and (D) solutions are, if there are, close to each other. It can be seen that the eigenvalue of mode $m (m=2,3,4)$ is, if there is, approximately ten times greater than the eigenvalue of mode ($m-1$) for the Hunter (B), (C) and (D) solutions.

\begin{figure}[htb]
\centerline{
\epsfxsize 18cm \epsfysize 20cm
\epsfbox{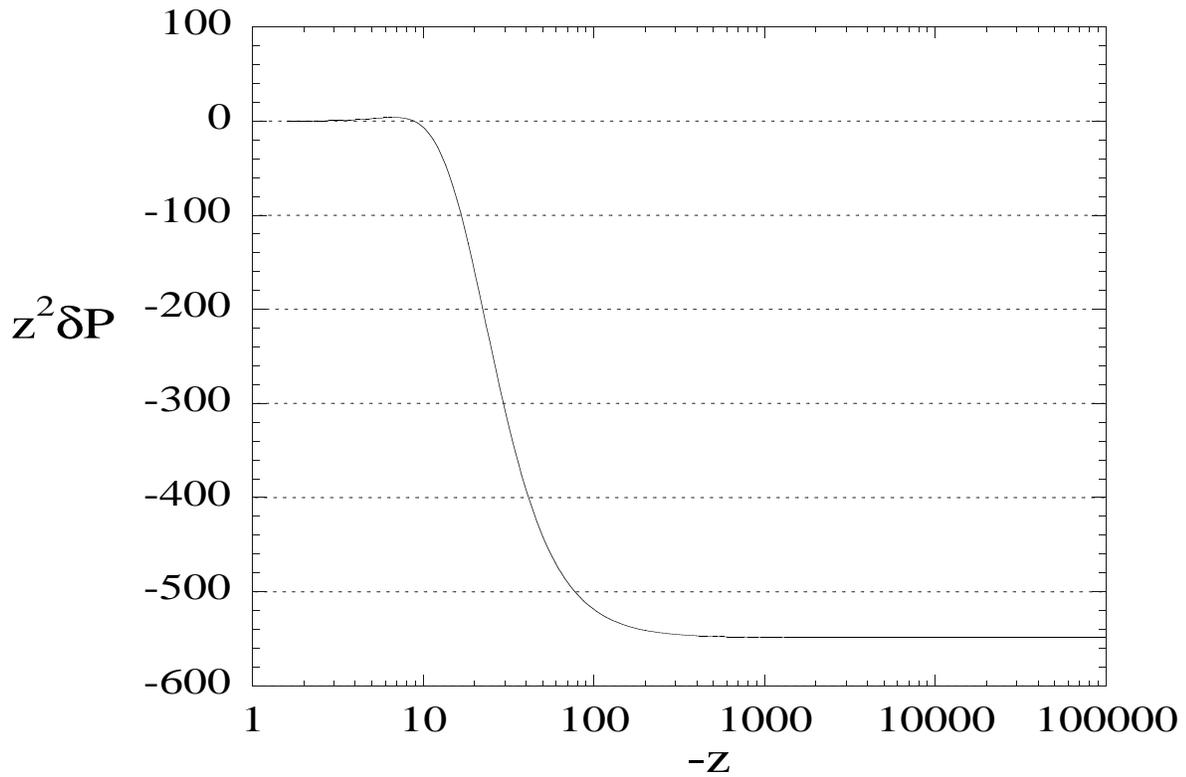}}
\caption{An unstable mode function of 
the density perturbation $z^2 \delta P=4\pi G t^2 \delta \rho$ for the Hunter (A) solution is plotted. A node can be seen in this mode function. It has a large amplitude near the center and a small amplitude with the 
opposite sign near the sonic point. This perturbation makes the 
concentration of the density strong or weak.}
\label{HAmodez2P}
\end{figure}

\begin{figure}[htb]
\centerline{
\epsfxsize 18cm \epsfysize 20cm
\epsfbox{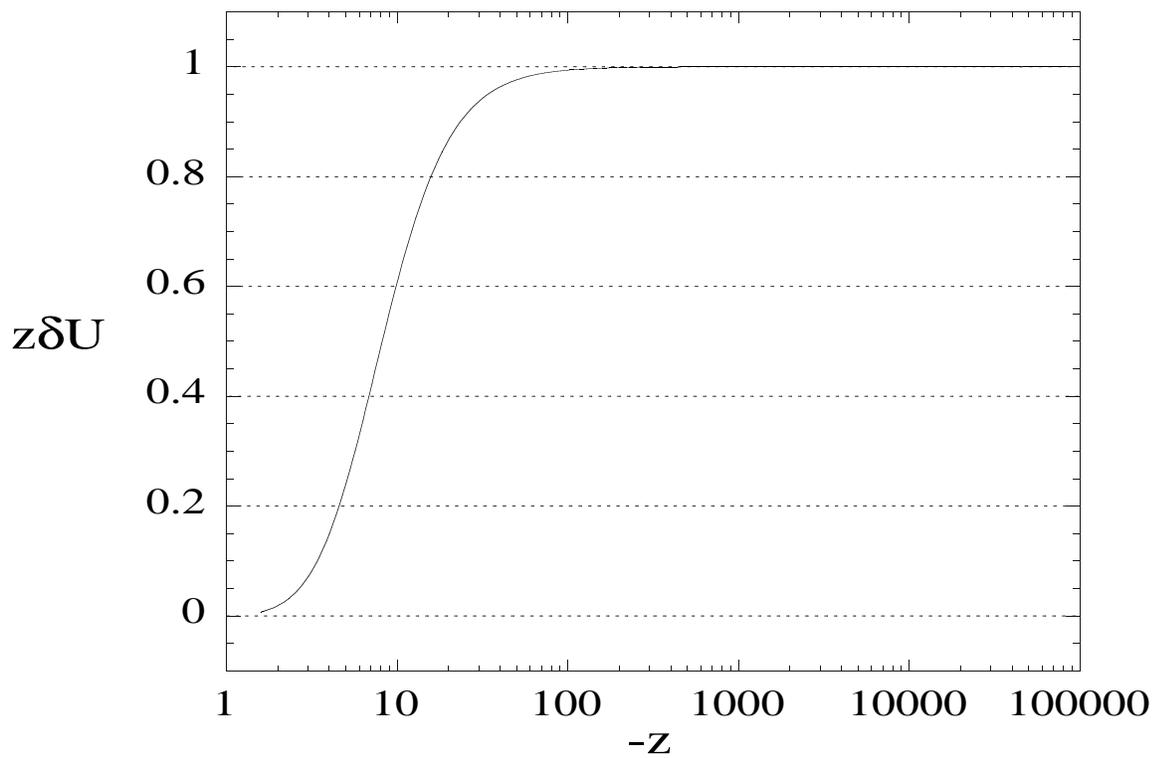}}
\caption{An unstable mode function of the 
velocity perturbation $z \delta U=-z \delta v/c_s$ 
for the Hunter (A) solution is plotted. No node can be seen in this mode function. It is found that the ``positive'' 
perturbation enhances the collapse of the gas spheres,
while the ``negative'' one promotes the gas to disperse away.}
\label{HAmodezU}
\end{figure}

The mode functions of the perturbations for the Hunter (A) solution are shown in Figs. \ref{HAmodez2P} and \ref{HAmodezU}. The mode function of the density perturbation has a node.  It has a large amplitude near the center and a small amplitude with the 
opposite sign near the sonic point. This perturbation makes the 
concentration of the density strong or weak. The mode function of the velocity perturbation does not have a node. 
It is found that the ``positive'' 
perturbation enhances the collapse of the gas spheres,
while the ``negative'' one promotes the gas to disperse away.

\section{DISCUSSIONS}

We have investigated the stability of the Larson-Penston solution, 
the homogeneous collapse solution and the 
Hunter (A)-(D) solutions by a normal mode analysis, 
assuming that the eigenvalue is positive. 
This assumption can be justified 
for $\Re \sigma >1$ (See Appendix in~\cite{hn1997}). 
In addition, the results of the Lyapunov 
analysis by~\cite{kha1999} for $c_{s}\ll c$, where $c$ is the speed of light, 
strongly suggests the validity of this assumption
at least for the Hunter (A) solution.

From the results in the previous section, 
the Hunter (A)-(D) solutions are unstable and not likely 
to be realized. It has been shown that both the homogeneous 
collapse solution and the Larson-Penston solution have no unstable mode. 
Since the homogeneous collapse solution has kink instability as 
proved by Ori and Piran~\cite{op1988}, the Larson-Penston solution 
is the best description of the central part of generic spherical collapse of isothermal gas. This result is consistent with the numerical 
simulations~\cite{hunter1977,ti1999,fc1993}.  

The eigenvalues of the Hunter (B) and (D) solutions 
obtained here are slightly different from the result of Hanawa and Nakayama~\cite{hn1997}, however the eigenvalues of the Hunter (A) and (B) solutions obtained here are consistent with those of the Newtonian limit in the general relativistic analysis~\cite{maison1996,hm2001,kha1999}. We believe that the eigenvalues obtained in the present paper are 
quite accurate.

Whitworth and Summers claimed that the Hunter (A) solution 
(and also the Hunter (C) solution) is unacceptable since 
it cannot pass through the second sonic point regularly 
(see Figs. \ref{U} and \ref{P})~\cite{WS1985}. However, we note 
that we can prepare regular initial density profile which 
develops these solutions. The fact that these solutions cannot pass
thorough the second sonic point analytically only implies that 
the self-similarity and regularity requirements are incompatible in the evolution of the sphere for $t>0$ and does not rule out
the possibility that they may describe generic 
collapse of the isothermal sphere leading to the core formation.
Only by the results of the present analysis, we can deny this 
possibility. 

We have found that the Hunter (A) solution has a single unstable mode.
From the discussions in use of a 
renormalization group~\cite{kha1995},
it implies that this solution is a critical solution
of some critical phenomena.
The obtained eigenvalue of the unstable mode 
for the Hunter (A) solution is $9.4637$, from which the critical exponent is calculated as 
$\gamma=\sigma^{-1}\simeq 0.10567$. 
The order parameter in this Newtonian case 
will be the collapsed mass in place of the formed 
black hole mass in general relativity.
It is clear that full numerical simulations will give a complete picture of 
the Newtonian critical behavior. 

\section{SUMMARY}

It is 
shown by means of a normal mode analysis that the Hunter (A)-(D) solutions are 
unstable, while
neither the Larson-Penston 
solution nor the homogeneous collapse 
one have an analytic unstable mode.
Since the homogeneous collapse solution is known 
to suffer the kink instability,
this result and recent numerical simulations strongly support 
a proposition that the Larson-Penston solution will be realized in 
astrophysical situations. 
It is also shown that the Hunter (A) solution has a single unstable mode,
which implies that it is a critical solution associated with 
some critical phenomena
which are analogous to those in general relativity.
The critical exponent $\gamma$ is calculated as $\gamma\simeq 0.10567$.  
In contrast to the general relativistic case, the order parameter will be 
the collapsed mass.

\acknowledgments

We are grateful to T.~Hanawa and S.~Inutsuka for useful comments and J.~Overduin for his critical reading of our paper.
We would also like to thank K.~Maeda for continuous encouragement. 
This work was partly supported by the 
Grant-in-Aid for Scientific Research (No. 05540)
from the Japanese Ministry of
Education, Culture, Sports, Science and Technology.

\appendix
\section{On the perturbations of the homogeneous collapse solution}

Changing the position variables from $r$ to $x$ as

\begin{eqnarray}
r= a(t) x,
\end{eqnarray}
the time derivative at fixed $r$ of a function $f=f(t,x=r/a)$ is

\begin{eqnarray}
\left(\frac{\partial f}{\partial t}\right)_r=\left(\frac{\partial f}{\partial t}\right)_x-\frac{\dot a}{a}x \left(\frac{\partial f}{\partial x}\right)_t,
\end{eqnarray}
where the gradient with respect to $x$ at fixed time is

\begin{eqnarray}
\left(\frac{\partial f}{\partial r}\right)_t=\frac{1}{a}\left(\frac{\partial f}{\partial x}\right)_t.
\end{eqnarray}
Then the equations (\ref{b1})-(\ref{b4}) become

\begin{eqnarray}
&&\frac{\partial\rho}{\partial t}-x\frac{\dot a}{a}\frac{\partial \rho}{\partial x}+\frac{1}{ax^2}\frac{\partial}{\partial x}(x^2 \rho v)=0,\label{p1} \\
&&\frac{\partial v}{\partial t}+\frac{v-{\dot a}x}{a}\frac{\partial v}{\partial x}+\frac{c_s^2}{\rho a}\frac{\partial\rho}{\partial x}+\frac{G M}{a^2 x^2}=0, \\
&&\frac{\partial M}{\partial t}+\frac{v-{\dot a}x}{a}\frac{\partial M}{\partial x}=0, \\
&&\frac{1}{a}\frac{\partial M}{\partial x}=4\pi a^2 x^2 \rho. \label{p4}
\end{eqnarray}
We will write the perturbations of the homogeneous background as

\begin{eqnarray}
v&=& {\dot a} x+ u(x,t), \\
\rho&=&\rho_b(t)[1+\delta(x,t)], \\
M&=&\frac43 \pi \rho_b a^3 x^3+ \Delta(x,t), 
\end{eqnarray}
where the expansion factor $a(t)$ and the mass density of the homogeneous background $\rho_b(t)$ satisfy

\begin{eqnarray}
a=a_0 t^{\frac23}, \quad \rho_b =\frac{1}{6\pi G t^2},
\end{eqnarray}
where $a_0$ is a constant. We obtain the time evolution equation for the mass density contrast from the equations (\ref{p1})-(\ref{p4}) in the linear perturbation theory (see~\cite{peebles} P.116) ,

\begin{eqnarray}
\frac{\partial^2 \delta}{\partial t^2}+ 2\frac{\dot a}{a}\frac{\partial \delta}{\partial t}=4\pi G \rho_b \delta+\frac{c_s^2}{a^2}\frac{1}{x^2}\frac{\partial}{\partial x}\left(x^2 \frac{\partial \delta}{\partial x}\right). \label{apa9}
\end{eqnarray}
We assume the time dependence of the perturbation as (\ref{tdepend}), 

\begin{eqnarray}
\delta=t^{-\sigma} e^{A(z)}.
\end{eqnarray}
In the homogeneous model, (\ref{apa9}) becomes

\begin{eqnarray}
&&\left[\frac{\sigma}{t^2} +\frac{z^2}{t^2}A''+\left(\frac{\sigma}{t}+\frac{z}{t}A'\right)^2\right]+\frac{4}{3t}\left(\frac{\sigma}{t}+\frac{z}{t}A'\right) \nonumber \\
&& =\frac{2}{3t^2}+\frac{c_s^2 z^2}{a_0^2 t^{\frac43}}A''.
\end{eqnarray}
The above equations have two and only two eigenvalues $\sigma=1$ or $-2/3$. $\sigma=-2/3$ corresoponds to a stable mode while $\sigma=1$ corresoponds to a 
ghost mode.

\section{Ghost mode}

It is found that the system has a ghost mode, $\sigma=1$. The mode functions are given by

\begin{eqnarray}
&&P_1=P' e^{\tau}, \\
&&U_1=U' e^{\tau}. 
\end{eqnarray}
This mode corresponds to the following transformation:

\begin{eqnarray}
&&t \to t- \varepsilon, \\
&&r \to r,
\end{eqnarray}
or, equivalently, 

\begin{eqnarray}
&&\tau \to \tau-\varepsilon e^{\tau}, \\
&&z \to z+\varepsilon e^{\tau} z.
\end{eqnarray}
This mode has no physical meaning for stability.

\newpage

\end{document}